\documentclass[twocolumn,showpacs,floatfix,aps,prb]{revtex4}
\usepackage{graphicx}
\usepackage{units}
\usepackage{bm}
\usepackage{amsmath}
\usepackage{amsfonts}
\usepackage{amssymb}
\usepackage[thinspace,squaren,pstricks,derivedinbase,derived]{SIunits}
\usepackage{CJKutf8}
\usepackage[pdfpagelabels, colorlinks, citecolor=blue]{hyperref}

\begin{document}

\title{On the phonon dispersion relation of single-crystalline $\beta$--FeSe}

\author{Khalil Zakeri}
\email{khalil.zakeri@partner.kit.edu}
\author{Tobias Engelhardt} \affiliation {Heisenberg Spin-dynamics Group, Physikalisches Institut, Karlsruhe Institute of Technology, Wolfgang-Gaede-Str. 1, D-76131 Karlsruhe, Germany}
\author{Thomas Wolf}
\author{Matthieu Le Tacon}
\affiliation{Institut f\"ur Festk\"orperphysik, Karlsruhe Institute of Technology, Hermann-v.-Helmholtz-Platz 1, D-76344 Eggenstein-Leopoldshafen, Germany}

\begin{abstract}
We report on the phonon spectrum probed at the $\beta$--FeSe(001) surface by means of high-resolution electron energy-loss spectroscopy (HREELS). Single crystals of $\beta$--FeSe are cleaved under ultra-high vacuum conditions and are subsequently measured below and above the nematic transition temperature. In total we observe five phonon modes and a phonon cutoff energy of about 40 meV. We identify the origin of each phonon mode based on the selection rules of HREELS and by comparing the experimental results to the ones of \emph{ab initio} density functional calculations. The most prominent phonon modes $A_{1g}$, $B_{1g}$, and $A_{2u}$  appear at energies of about $20.5$ and $25.6$  and $40$ meV, respectively. These phonon modes disperse rather weakly while changing the momentum from zero up to the zone boundary, indicating that they are mainly of optical nature. A comparison between our results and the results of \emph{ab initio} calculations indicates that there must be a mutual interplay between magnetism and lattice dynamics in this compound, similar to the other Fe-based superconductors. Finally, we comment on the role of temperature on the phonon modes probed at the $\bar{\Gamma}$--point. It is observed that both the $B_{1g}$ and $A_{2u}$ phonon modes undergo a downward shift while increasing the temperature from 15 to 300 K. In the case of the $A_{2u}$ mode this shift is about 1.5 meV.
\end{abstract}

\maketitle

\section{Introduction}

Among all Fe-based superconductors $\beta$--FeSe is, structurally, the simplest one and has been investigated extensively in past years \cite{Hsu2008}. The superconducting transition temperature in this material is $T_c \simeq 8$ K at ambient pressure  \cite{Hsu2008} and can increase to  $T_c \simeq 37$ K at high pressures \cite{Medvedev2009}. Similar to the FeAs planes in LiFeAs, LaFeAsO and BaFe$_2$As$_2$, which are prototypes of the well-known families of Fe-As based high-$T_c$ superconductors,  $\beta$--FeSe contains also square planar sheets of tetrahedrally coordinated Fe atoms \cite{Johnson2015}. The recent discovery of superconductivity up to 100 K in the single unit cell of FeSe grown on SrTiO$_3$ has brought this material to the centre of attention and to the forefront of research in the field of superconductivity \cite{Ge2015}.

Although there are distinct differences between FeSe and the other Fe-based superconductors, it is now generally accepted that similar to the other Fe-based compounds, FeSe also belongs to the family of unconventional superconductors, of which their superconductivity is not directly mediated by the electron-phonon coupling  \cite{Stewart2011, Lumsden2010, Paglione2010, Johnson2015, Ewings2008, Kotegawa2012, Inosov2015, Rahn2015, Park2011}. Recent inelastic neutron scattering experiments have revealed the so-called spin resonance mode, suggesting that the superconductivity in this material is associated with spin excitations \cite{Ma2017,Wang2015}. However, observation of the isotope effect has opened up many questions regarding the role of phonons in the superconductivity \cite{Khasanov2010}. Although the material appears simple at first glance, the actual
role of the structural degrees of freedom and lattice dynamics in superconductivity remains unresolved \cite{Gnezdilov2013, Baek2015, Chubukov2015, Mazin2015}. Many efforts have been devoted to investigate the phonons in this material \cite{Phelan2009, Ksenofontov2010, Xia2009, Okazaki2011, Litvinchuk2008, Kumar2010, Um2012, Gnezdilov2013}. However, due to the lack of large single crystals, most of those experiments are performed on polycrystalline or powder samples. Only a few experiments are reported on the single-crystalline samples \cite{Xia2009, Okazaki2011, Litvinchuk2008, Kumar2010, Gnezdilov2013}, but those experiments focus only on the phonon modes at the Brillouin zone center and mainly on the Raman active modes.

Here we aim to look at the problem from the perspective of surface science. We therefore prepare clean and well-ordered $\beta$--FeSe(001) surfaces and investigate them under ultra-high vacuum conditions. We provide the first phonon spectrum recorded on the surface of $\beta$--FeSe(001) single crystals by means of electrons, with the particular attention on the $A_{1g}$, $B_{1g}$ and $A_{2u}$ phonon modes, observed at the zone center. Our results indicate that these phonon modes originate from the atomic displacements of either Se or Fe atoms along the $c$-axis, normal to the surface and hence can be very efficiently excited by electrons. The modes show a rather weak dispersion, as expected.  By comparing our results to those of available \emph{ab initio} density functional calculations, we discuss the interplay between magnetism and the lattice dynamics in this compound.  In addition, we comment on the role of temperature on the observed phonon modes.

\section{Experimental details}
\begin{figure}[t]
\vspace{12pt} \center
\resizebox*{0.8\columnwidth}{!}{\includegraphics{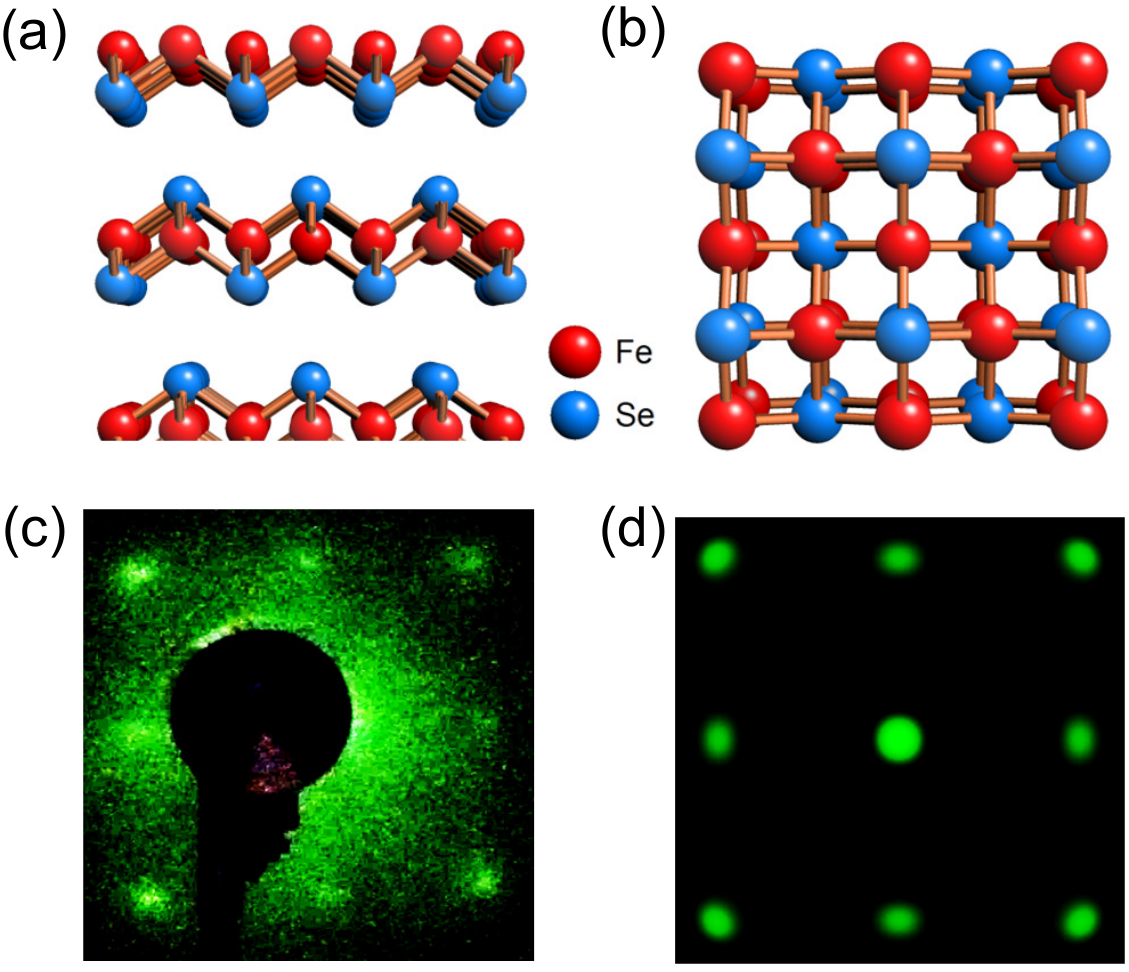}}
\caption{(color online) (a) The crystal structure of $\beta$-FeSe with a cut along the $a-b$ plane. (b) The (001) surface of $\beta$-FeSe. The Fe atoms are represented by the red balls and the Se atoms are represented by the blue balls. (c) The LEED pattern taken on the $\beta$-FeSe(001) surface at an electron energy of 92 eV and at room temperature. The simulated LEED pattern of the same surface and at the same energy.}
\label{Fig1}
\end{figure}

High-quality single crystals of $\beta$-FeSe were synthesized from Fe and Se powders mixed in an atomic ratio 1.1 to 1 and sealed in an evacuated SiO$_2$ ampoule together with an eutectic mixture of KCl and AlCl$_3$. Details of the crystal growth method may be found elsewhere (see for example Ref. [\onlinecite{Boehmer2013}]).
Samples were fixed on a holder and were transferred into the ultra-high vacuum (UHV) chamber. They were cleaved at a pressure of about $P\thickapprox 1 \times 10^{-10}$ mbar at room temperature. This leads to a clean and well-ordered FeSe(001) surface, with a ($1\times1$) surface reconstruction. In Fig. \ref{Fig1}(a) the crystal structure of $\beta$-FeSe is shown with a cut along the $a-c$ plane. The Fe atoms are covalently coordinated with Se anions above and below the Fe plane. The FeSe sheets are weakly connected with a weak van der Waals interaction along the $c$ axis and hence the crystal can easily be cleaved. The resulting surface is the (001) surface, parallel to the $a-b$ plane and perpendicular to the $c$ axis. This surface is schematically shown in Fig. \ref{Fig1} (b). The structural analysis of the surface was performed by means of low-energy electron diffraction (LEED) at room temperature.  A typical LEED pattern recorded at an electron energy of 92 eV and the corresponding simulated pattern are shown in Figs. \ref{Fig1} (c) and (d), respectively. The clear ($1\times1$) LEED pattern indicates a contamination free surface with a low step density.

The phonon spectrum of the (001) surface was measured by means of our spin-resolved high-resolution electron energy loss spectrometer \cite{Zakeri2014}. The incident electron energy was set on $4.076$ eV. The full width at half  maximum  (FWHM) of the elastic peak was about 6.3 meV. Note that the actual resolution of the spectrometer is even higher than this value. The additional broadening might be due to different reasons. For instance the presence of low energy phonons near the elastic peak can lead to a broadening of this peak. Likewise the presence of the twins and grain boundaries of the crystal can also broaden this peak. Such twins and grain boundaries lead to the broadening of the LEED spots shown in Fig. \ref{Fig1} (c). The scattering plane was chosen to be along the [110]-direction. Based on the scattering selection rules, this would allow probing all the possible phonon modes which have a polarization along any of the main crystallographic directions (all phonons with polarization along either the [100]- or the [010]-direction can be probed) \cite{Ibach1982}. This means that both even and odd modes of the (100) plane are allowed to be excited.  In this geometry the phonon dispersion relation is probed along the $\overline{\Gamma}$--$\overline{\rm M}$ direction of the surface Brillouin zone.
The total scattering angle (the angle between the incident and scattered beam) was set on $80^{\circ}$. This means the angle of incident and scattered beam with respect to the surface normal at the specular reflection was $40^{\circ}$. For probing the phonon dispersion relation the energy-loss spectra were recorded at different in-plane wave vector transfers $\Delta k_{\parallel}=k_{f\parallel}-k_{i\parallel}$. Here $k_{f\parallel}$ and $k_{i\parallel}$ are the in-plane wave vectors of incident and scattered beam, respectively. Different values of $\Delta k_{\parallel}$ were achieved by changing the angles of the incident and scattered beam with respect to the surface normal. A spin polarized electron beam was used for the experiments. However, phonons are spin independent collective excitations and therefore no spin dependence was observed, as expected. Thus the spectra were recorded in the so-called spin integrated mode. In this mode both the spin-flip and non-spin-flip contributions to the scattering intensity are present.

\section{Results and Discussions}

\begin{figure}[h]
\vspace{12pt} \center
\resizebox*{0.99\columnwidth}{!}{\includegraphics{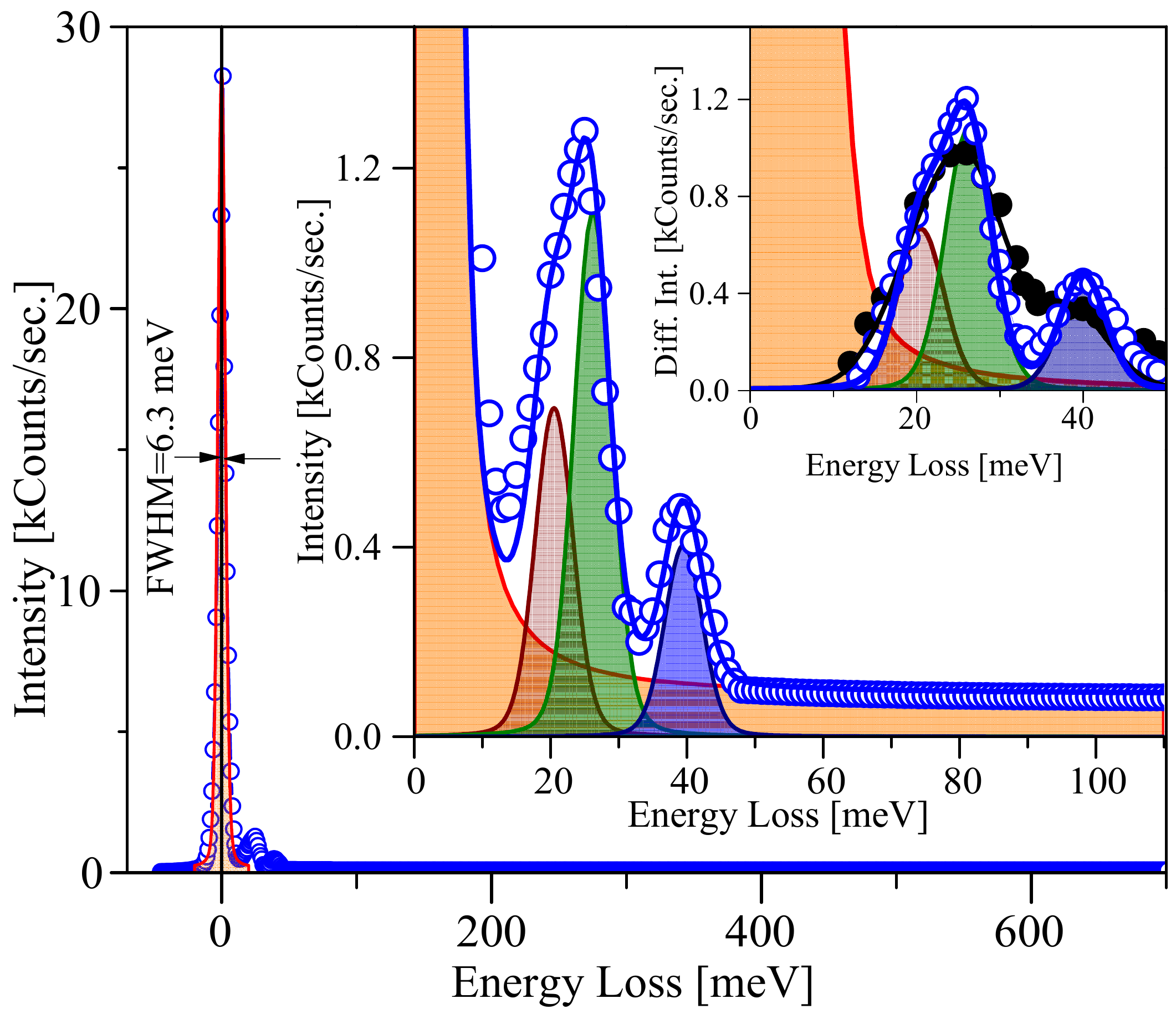}}
\caption{(color online) High-resolution loss spectrum recorded on the $\beta$-FeSe(001) surface at 15 K. The spectrum is recorded with an incident energy of $4.076$ eV at the specular geometry. Insets shows a magnified part of the spectrum where different phonon modes are observed. The upper right inset shows the experimental data after subtraction of the elastic line. The data are fitted by a convolution of Voigt functions. The data recorded at $T=300$ K are shown by solid black circles for a comparison.}
\label{Fig2}
\end{figure}

A typical spin-integrated  spectrum recorded at a sample temperature of about 15 K and at the specular geometry (with no parallel momentum transfer $\Delta k_{\parallel}=0$)  is shown in Fig. \ref{Fig2}. The spectrum is dominated by the presence of the elastic peak at the energy loss of zero.  Beside the elastic line one clearly observes peaks associated with the inelastic scattering of the electrons by different phonon modes.

In this geometry one clearly observes three different phonon modes. In order to extract the energies of these phonon modes from the experimental data, the spectra were evaluated by a set of Voigt functions. The same analysis was performed on the spectra with the elastic peak subtracted (right upper inset of Fig. \ref{Fig2}). We first fitted the quasi-elastic peak at the energy loss of zero and then subtracted it from the experimental data. With this only the loss intensities caused by the scattering of electrons from the vibrational excitations remain.  The remaining intensity profile was fitted by a set of Voigt profiles. We note that analyses of data with and without the quasi-elastic peak subtracted lead to exactly the same results (see Fig. \ref{Fig2}).  In the Voigt profiles the Gaussian broadening was chosen to be the experimental energy resolution (FWHM = 6.3 meV). The Lorentzian linewidth in such a case denotes the intrinsic phonon lifetime.

\begin{figure}[t]
\vspace{12pt} \center
\resizebox*{0.9\columnwidth}{!}{\includegraphics{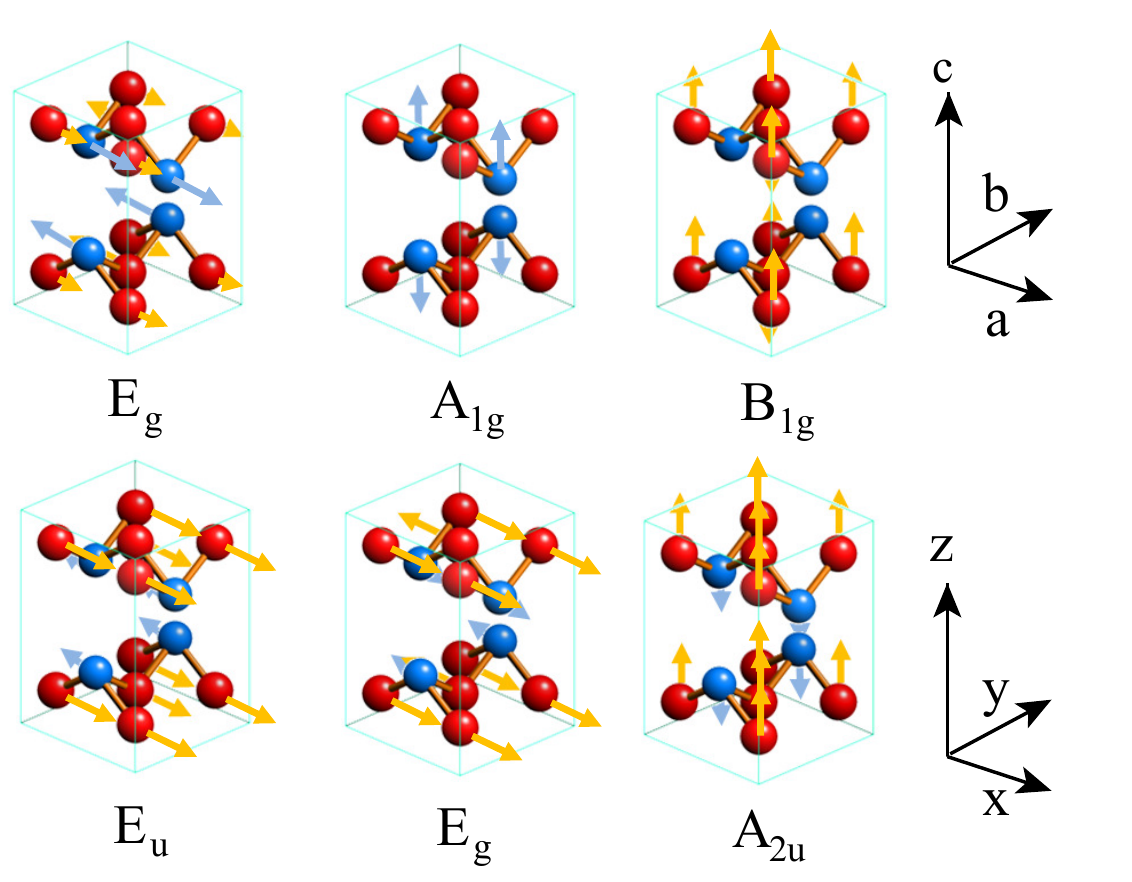}}
\caption{(color online)  All possible phonon modes of the $\beta$--FeSe with $P4/nmm$  symmetry. The Fe atoms are represented by the red balls while the Se atoms are represented by the blue balls. The displacement direction is represented by the small arrows on the Fe and Se sublattices.}
\label{Fig3}
\end{figure}

At room temperature FeSe has the tetragonal PbO structure and belongs to the space group $P4/nmm$. At low temperatures, below the nematic transition temperature, the crystal structure transforms into $Cmma$ \cite{Hsu2008}. Based on simple group theory consideration, one can predict the possible phonon modes of the system. The possible phonon modes of  $\beta$--FeSe are depicted in Fig. \ref{Fig3}.  Among the different phonon modes depicted in Fig. \ref{Fig3} the $A_{1g}$, $B_{1g}$ and $A_{2u}$ modes are the $z$ polarized modes, which involve vertical displacement of Se and Fe atoms. The $A_{1g}$ and $B_{1g}$ modes are the so-called  Raman active modes while $A_{2u}$  is an infrared active phonon. These modes are expected to be observed at the $\Gamma$--point of the three-dimensional bulk Brillouin zone \cite{Xia2009, Kumar2010, Okazaki2011, Gnezdilov2013, Hu2016}.

At the specular geometry  the parallel component $\Delta k_{\parallel}$ is zero. The kinematic model of electron scattering predicts that the intensity of a phonon that is polarized perpendicular to the surface should be much higher than the intensity of an in-plane polarized phonon, since the vertical component of the wave vector transfer $\Delta k_{\perp}$ is much larger than the parallel component $\Delta k_{\parallel}$ \cite{Ibach1982}. Hence mainly the phonon modes which have a polarization perpendicular to the surface shall be observed near the specular geometry $\Delta k_{\parallel} \sim 0$. The observed loss features at $\varepsilon \simeq20.5$, $25.6$ and $40$ meV can therefore be assigned to the $z$-polarized $A_{1g}$, $B_{1g}$ and $A_{2u}$ phonon modes, respectively.  Both the $A_{1g}$ and $B_{1g}$ modes have already been observed in the Raman scattering experiments \cite{Xia2009, Kumar2010, Okazaki2011, Gnezdilov2013, Hu2016}. Since the $A_{2u}$ mode is not a Raman active mode it has not been observed in those experiments.

As a side note we would like to point out that the typical probing depth of HREELS is only a few atomic layers below the surface layer. Generally the $A_{1g}$, $B_{1g}$, and $A_{2u}$ modes are referred to as the modes which are excited at the $\Gamma$--point of the three-dimensional bulk Brillouin zone. In the HREELS experiments the momentum in the direction perpendicular to the surface is not conserved and hence only the excitations in the two-dimensional surface Brillouin zone are probed. The modes observed in the two-dimensional surface Brillouin zone are the projected states of the three-dimensional Brillouin zone. Owing to the layered nature of FeSe the $A_{1g}$, $B_{1g}$, and $A_{2u}$ modes do not show any particular momentum dependence along the $\Gamma$--Z direction of the three-dimensional Brillouin zone (along this direction the phonon branches are rather flat). Hence their projection to the surface Brillouin zone would not result in any additional dispersion, in particular when projecting $\Gamma$ to $\overline{\Gamma}$.  Therefore the notations $A_{1g}$, $B_{1g}$, and $A_{2u}$ can also be used for the zone center of the two-dimensional surface Brillouin, $\overline{\Gamma}$--point.

\begin{figure}[t]
\vspace{12pt} \center
\resizebox*{0.90\columnwidth}{!}{\includegraphics{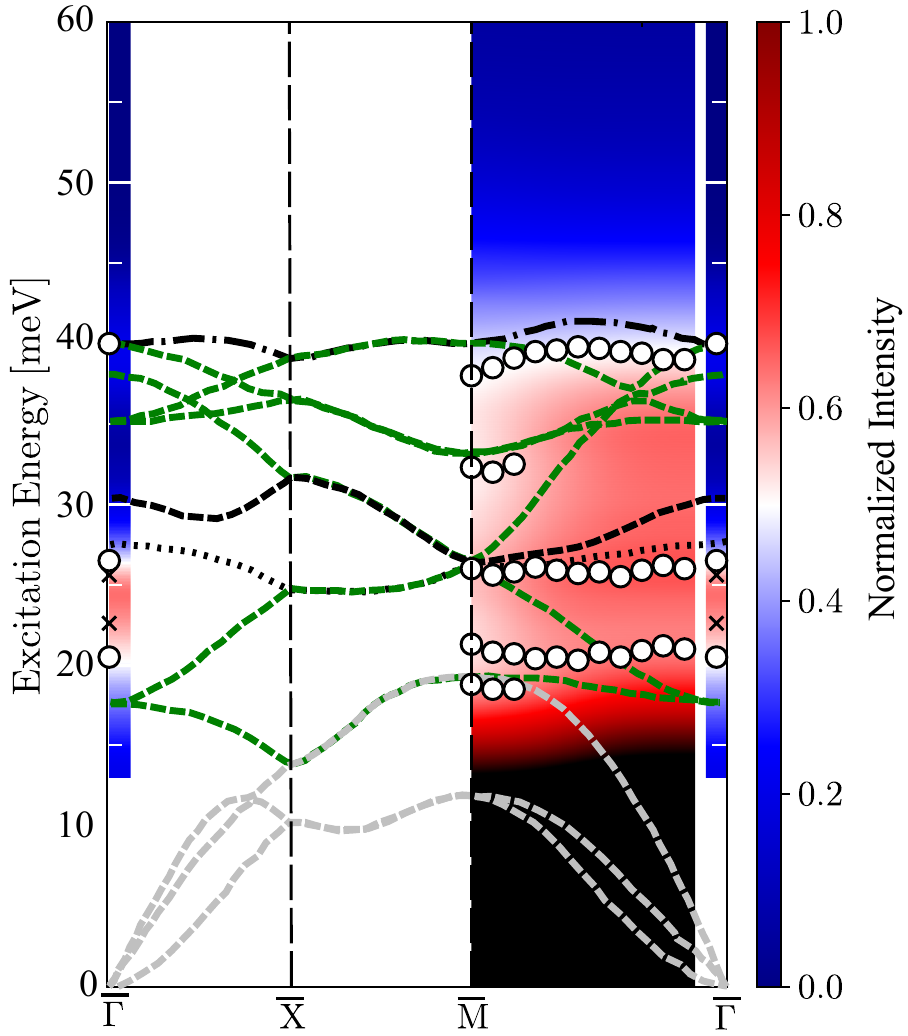}}
\caption{(color online) Phonon dispersion relation probed on the $\beta$-FeSe(001) surface at 15 K. The dispersion relation is probed along the $\overline{\Gamma}$--$\overline{\rm M}$ direction of the surface Brillouin zone. The color bar represents the scattering intensity. The spectra are normalized to the quasi-elastic peak. The open circles represent the peak positions (phonon energies). The calculated phonon dispersion relation for the FeSe in the nonmagnetic phase is also shown for a comparison (dashed and dotted curves). The black dotted (dashed) curve shows the dispersion of the $A_{1g}$ ($B_{1g}$) mode. The dashed-dotted curve indicates the dispersion relation of the $A_{2u}$ mode. The acoustic phonon modes are shown by the light gray color. The data are taken from Ref. \cite{Subedi2008}. The Raman modes measured by Gnezdilov et al. \cite{Gnezdilov2013} are also shown as crosses.}
\label{Fig4}
\end{figure}

In order to obtain the dispersion relation of the observed phonon modes, the spectra were recorded at different values of $\Delta k_{\parallel}$ and the results are presented in Fig. \ref{Fig4}. The data were recorded along the [110]-direction of the surface, which corresponds to the high-symmetry $\overline{\Gamma}$--$\overline{\rm{M}}$ line of the surface Brillouin zone. As is apparent from Fig. \ref{Fig4},  all the three $A_{1g}$,  $B_{1g}$ and $A_{2u}$ modes observed at the $\overline{\Gamma}$-point could also be observed at the higher wave vector transfers. The modes show a rather weak dispersion. At higher wave vector transfers, close to the $\overline{\rm M}$--point, two additional sets of excitations could clearly be observed. These excitations must exhibit  a large in-plane polarization and can therefore be assigned to those phonon modes which exhibit a large in-plane polarization e.g. $E_u$ and $E_g$ modes (see the discussion below).

\begin{table*}{t}
\caption{\label{tab:phononmodes}The energies of different phonon modes observed at the high symmetry points of the Brillouin zone. The experimental results recorded at the $\beta$--FeSe(001) surface are compared to the results of \emph{ab initio} calculations and also to the other experimental results obtained on similar systems. The values are given in meV. The abbreviations are defined as follows. SC: single-crystalline, PC: poly-crystalline, $\mu$C: microcrystals, NM: nonmagnetic, SAFM: stripe antiferromagnetic, CBAFM: checkerboard antiferromagnetic.}
\begin{ruledtabular}
\begin{tabular}{cccccccc}
 &\multicolumn{3}{c}{ }&\multicolumn{3}{c}{at $\overline{\Gamma}$--point\footnote{This point is equivalent to the $\Gamma$--point of the bulk Brillouin zone.}} &\multicolumn{1}{c}{near $\rm \overline{M}$--point\footnote{This point is equivalent to the $\rm M$--point of the bulk Brillouin zone.} }\\

 & &  & T [K]  &  $A_{1g}$ & $B_{1g}$ &  $A_{2u}$ &  $E_{g}/E_{u}$\\
\hline\hline
&This work & SC  & $15$  & $ 20.5$ & $25.6$ & $  40$  & $  32$\\
& Expt. \cite{Phelan2009}  & PC    & $1.5$ & $ 20.5$ & $24.5$ & $  39$ & $  31.5$\\
& Expt.\cite{Ksenofontov2010}  & PC & $ 10$ & $ 20.6$ & $25.5$ & $  38.7$ & $  31.5$\\
& Expt. \cite{Gnezdilov2013}  & SC & $7$   & $ 22.5$ & $25.6$ & $ -$& $ -$\\
FeSe& Expt. \cite{Kumar2010}  & SC  & 3   & $20$ & $28$ & $ -$& $ -$\\
& Calc. \cite{Subedi2008}  & NM  & $ -$   & $27.5$ & $30.5$ & $ 42$ & $ 35$\\
& Calc. \cite{Okazaki2011}  & NM & $ -$   & $22$ & $27.5$ & $ 36$ & $ -$\\
& Calc. \cite{Wang2012}  & NM  & $ -$   & $27.9$ & $29.7$ & $ 41.2$ & $ 30$--$35$\\
& Calc. \cite{Wang2012}  & SAFM  & $ -$    & $21.5$ & $27.8$ & $ 40.4$ & $ 30$--$35$\\
& Calc. \cite{Ye2013}  & CBAFM & $ -$  & $24.1$ & $26.6$ & $ 41$& $-$\\
\hline
FeTe$_{0.92}$& Expt.\cite{Xia2009}  & SC & $ 300$    & $ 19.7$ & $24.3$ & $ -$ & $ -$  \\
Fe$_{1.074}$Te & Expt.\cite{Okazaki2011}  & SC & $ 5$    & $ 19.6$ & $25$ & $ -$ & $ -$  \\
FeTe$_{0.6}$Se$_{0.4}$& Expt.\cite{Okazaki2011}  & SC & $ 5$    & $ 20$ & $25$ & $ -$ & $ -$  \\
Fe$_{1.02}$Te & Expt.\cite{Um2012}  & SC & $ 5$    & $ 19.7$ & $24.9$ & $ -$ & $ -$  \\
FeTe$_{0.78}$Se$_{0.22}$ & Expt.\cite{Um2012}  & SC & $ 5$    & $ 19.2$ & $25$ & $ -$ & $ -$  \\
Fe$_{0.95}$Te$_{0.56}$Se$_{0.44}$ & Expt.\cite{Um2012}  & SC & $ 5$    & $ 20$ & $25.4$ & $ -$ & $ -$  \\
\hline
(K,Sr)Fe$_2$As$_2$& Expt.\cite{Litvinchuk2008}  & $\mu$C  & $ 20$    & $22.8$ & $26$ & $ -$& $ -$\\
\end{tabular}
\end{ruledtabular}
\end{table*}

The phonon dispersion relation of $\beta$--FeSe has been calculated by different density functional based \emph{ab initio} methods \cite{Subedi2008, Nakamura2009, Kumar2010a, Wang2012}. We note that the calculations performed by different groups using different functionals have resulted a similar phonon dispersion relation (the shape of the dispersion relation in all the calculations is nearly the same). The absolute values of the energies are slightly different. A comparison between different results can be found in Tab. \ref{tab:phononmodes}. As an example the phonon dispersion relation calculated within the local-density approximation and the general potential linearized augmented plane-wave method, by Subedi et al. \cite{Subedi2008} is shown in Fig. \ref{Fig4} for a comparison. One immediately notices that these calculations which are performed for the nonmagnetic $\beta$--FeSe cannot precisely account for the phonon frequencies of  the $A_{1g}$,  $B_{1g}$ and $A_{2u}$ modes. For example according to the results of calculations at the zone center the $A_{1g}$ and $B_{1g}$ modes should have energies of about 27.5 meV and 30.4meV, respectively. They should merge together in the midway of $\bar{\Gamma}$ to $\bar{\rm M}$. These modes are shown  in Fig. \ref{Fig4} by black dotted and dashed lines, respectively. However our experiments show that these modes possess energies of about $20.5$ and $25.6$ meV, respectively and do not disperse with wave vector. A comparison between our results and the ones reported in the literature on FeSe and the other similar systems is provided in Tab. \ref{tab:phononmodes}. Our results near the $\bar{\Gamma}$--point are close to the values reported by Gnezdilov et al. \cite{Gnezdilov2013} obtained on single-crystalline samples by means of Raman scattering and obtained by Ksenofontov et al. \cite{Ksenofontov2010} on polycrystalline samples using $^{57}$Fe nuclear inelastic scattering. We note that the energy of the $B_{1g}$ phonon (25.6 meV) is in excellent agreement with the previous report on the bulk single crystals. On the other hand, the value of the $A_{1g}$ mode at the $\overline{\Gamma}$--point is smaller by 10\% when comparing our results to the ones of the single-crystalline bulk samples \cite{Gnezdilov2013}. The $A_{1g}$ is associated with the motion of the Se atoms. Since the crystal is cleaved so that the surface is Se terminated, one would expect that the Se atoms at the surface are subject to a weaker interatomic coupling. This would naturally lead to a lower energy of the surface phonons. We note that the Fe$_2$Se$_2$-layers interact rather weakly via the van der Waals interaction. This means that the surface effects are not as strong as what is observed in the case of materials with covalent, metallic or ionic bonding. A comparison between the results of calculations performed for nonmagnetic FeSe and for the stripe antiferromagnetic phase of FeSe shows that the $A_{1g}$ and $B_{1g}$  modes are sensitive to the magnetic state of the system \cite{Wang2012}.  These calculations result in the values of 21.5 and 27.8  meV for the $A_{1g}$ and $B_{1g}$ modes, respectively, when a stripe antiferromagnetic order is assumed as the ground state. Compared to the experimental value one recognizes that both modes are slightly overestimated, in particular the $B_{1g}$ mode.  Calculations performed by Ye et al. \cite{Ye2013} assuming a checkerboard antiferromagnetic ground state predict the values of 24.1 and 26.6  meV for the $A_{1g}$ and $B_{1g}$ modes, respectively. Also in this case the phonon energies are overestimated. However, now the $B_{1g}$ mode agrees better with the experiment and the $A_{1g}$  is overestimated by about 3.5 meV. A similar effect has been observed for the other Fe-based materials e.g iron-pnictides \cite{Yildirim2009} and other iron-chalcogenides \cite{Okazaki2011, Um2012}, indicating similarities between FeSe and the other Fe-based superconductors. It has been discussed that similar to other Fe-based superconductors, the magnetism has a considerable impact on the lattice dynamics of FeSe and determines the phonon frequencies \cite{Nakamura2009, Kumar2010a, Wang2012}. The effect is discussed in terms of spin--phonon coupling \cite{Gnezdilov2013, Kumar2010a, Wang2012, Ye2013}. In optical spectroscopy experiments spin-phonon coupling can induce peculiar features on the shape of the spectra, leading to an asymmetric lineshape \cite{Homes2016}. In such a case the excitation peak may be described by a Fano lineshape, instead of a Lorentzian. We did not, however, observe any additional effect on the shape of the spectra which could be associated with the spin-lattice coupling. This might be due to the fact that the underlying physical mechanism behind the excitation of phonons by means of electrons is entirely different from that of the polarized photons \cite{Ibach1982}.
Interestingly, the calculations for the nonmagnetic FeSe predict the $A_{2u}$ mode rather accurately. At the $\bar{\Gamma}$--point the agreement is perfect.  Away from the $\bar{\Gamma}$--point the measured dispersion relation of this mode is very similar to what has been predicted by theory (the black dashed-dotted curve in Fig. \ref{Fig4}), in spite of the small energy shift of about 2 meV. The experimentally probed dispersion relation of this mode exhibits a slight ``downward" parabolic shape as predicted by the calculations.

The phonon mode observed at about 32-33 meV in Fig. \ref{Fig4} is most likely caused by the $E_u$ or the high-energy $E_g$ mode (the one which involves a larger  displacement of the Fe atoms). As these two modes are very close in energy, it could be that the feature observed at about 32-33 meV is a combination of these two  modes. The mode at about 18-19 meV can be assigned to the low-energy $E_g$ mode (the one which involves the displacement of the Se atoms). It could also be the top of the highest energy acoustic phonons, as the energy of the acoustic phonon branches reaches the value of 18-19 meV near the $\bar{\rm M}$--point (see the dashed gray curves in Fig. \ref{Fig4}). Interestingly these modes are also accurately predicted by calculations, indicating the robustness of these modes with respect to the magnetic state of the system.

Finally, in order to investigate the effect of temperature on the observed phonon modes the same experiments were performed at room temperature. The data recorded at T=300 K are shown by black solid circles in the upper-right corner of Fig. \ref{Fig2} for a comparison. A small shift toward lower energies was observed for the $B_{1g}$  and $A_{2u}$ modes, while warming up the sample to room temperature. In addition the linewidth of both modes increases with temperature. It is not easy to quantify the temperature-induced shift of the $B_{1g}$ mode, as it moves toward the $A_{1g}$ mode and, at the same time, broadens. The $A_{2u}$ phonon mode undergoes a downward shift from 40 meV to about 38.5 meV. The shift is clearly visible in the data presented in Fig. \ref{Fig2}.  An unusual temperature dependence of the surface phonon modes has been observed in the case of Ba(Fe$_{1-x}$Co$_{x}$)$_2$As$_2$ \cite{Teng2013}. We did not, however, observe any additional effect associated with the surface properties.
The temperature-induced shift of the $A_{1g}$ and $B_{1g}$ Raman active modes has been observed in the Raman experiments by Gnezdilov et al. \cite{Gnezdilov2013}. It has been observed that the $B_{1g}$ mode undergoes a shift of about 1.5 meV, while the $A_{1g}$ mode exhibits a very small shift of about 0.35 meV. This behavior is in agreement with our observations. Such a behavior could not be explained only by the contraction of the lattice during cooling. An analysis based on the Gr\"{u}neisen law has shown that the $\gamma$ parameter must be about 1.04 for the $A_{1g}$  mode and about 5.09 in the case of the $B_{1g}$ mode \cite{Gnezdilov2013}. Both of these values deviate from the conventional value of about 2. We performed the same analysis for the $A_{2u}$ mode and obtained the value of the Gr\"{u}neisen parameter $\gamma\simeq 3.25 \pm 0.2$ for this mode. This value also deviates from the conventional value, meaning that the phonon hardening, observed during cooling, must be associated with the changes of the spin state of the Fe atoms within the FeSe unit cell. This is another indication of the presence of the spin-phonon coupling in the system.

The temperature dependence of  the $B_{1g}$ mode in the Fe$_{1+y}$Se$_{1-x}$Te$_x$ compound has been investigated \cite{Um2012}. In the case of Fe$_{1.02}$Te it has been observed that both the phonon frequency and the linewidth show a kink at the magnetic ordering temperature ($T_N$=67 K). The kink gradually evolves into a rather smooth behavior as Te atoms are substituted with Se. In the case of Fe$_{0.95}$Se$_{0.56}$Te$_{0.44}$  only a very smooth temperature dependence of phonon frequency and linewidth is observed.  This observation has been attributed to a peculiar coupling of the $B_{1g}$ phonon mode of the Fe$_{1+y}$Se$_{1-x}$Te$_x$ system to magnetic fluctuations \cite{Um2012}.

Recently the temperature dependence of the infrared active mode has been investigated for Fe$_{1.03}$Te and Fe$_{1.13}$Te as well as for the superconducting FeTe$_{0.55}$Se$_{0.45}$. It has been observed that these modes exhibit an asymmetric Fano lineshape which is superimposed on a background of Drude free carriers and symmetric Lorentz oscillators. This behavior has been discussed in terms of an anharmonic decay process \cite{Homes2016}.
In the case of the superconducting sample the infrared active $E_u$ mode has been observed to display an asymmetric lineshape at all temperatures, in particular in the temperature range of 100--200 K. This observation has been attributed to the presence of the spin-/electron-phonon coupling in the system\cite{Homes2016}.

\section{Conclusion}

The dispersion relation of different phonon modes probed at the $\beta$--FeSe(001) surface by means of HREELS is reported. Five different phonon modes and a phonon cutoff energy of about 40 meV are observed. The origin of each phonon mode is identified based on the selection rules of HREELS and in comparison with the results of \emph{ab initio} calculations. The $A_{1g}$, $B_{1g}$, and $A_{2u}$ phonons are found to possess energies of about $ \simeq20.5$ and $25.6$  and $40$ meV at the zone center ($\bar{\Gamma}$--point). All modes show a rather flat dispersion relation. The measured energies of these phonon modes do not match the values predicted by the available density functional calculations assuming a nonmagnetic \cite{Subedi2008}, a stripe antiferromagnetic \cite{Wang2012} or checkerboard antiferromagnetic \cite{Ye2013}  structure for $\beta$--FeSe, indicating a mutual interplay between magnetism and lattice dynamics in this system. The role of temperature on the phonon modes probed at the $\bar{\Gamma}$--point is investigated. A downward shift is observed for both the $B_{1g}$ and $A_{2u}$ phonon modes while increasing the temperature from 15 to 300 K. The effect seems to be largest on the $A_{2u}$ phonon mode which exhibits a shift from 40 meV to about 38.5 meV. Such a temperature-induced phonon softening must be due to the changes of the spin state of the Fe atoms within the FeSe unit cell. This is another indication of the presence of the spin-phonon coupling in the system.

\section{Acknowledgments}
This work has been supported by the Deutsche Forschungsgemeinschaft (DFG) through the Heisenberg Programme ZA 902/3-1.
\bibliographystyle {apsrev}
\bibliography {./FeSeRefs}

\end{document}